\documentclass[aps,pra,twocolumn,amsmath,amssymb,nofootinbib,showpacs,superscriptaddress]{revtex4-1}
\usepackage[english]{babel}
\usepackage{latexsym}
\usepackage{graphics}
\usepackage{graphicx}
\usepackage{epsfig}
\usepackage{color}
\usepackage{bm}
\usepackage{amsmath}
\usepackage{amssymb}
\usepackage{amsthm}
\usepackage{dcolumn}
\usepackage{bm}
\usepackage{float}
\usepackage{hyperref}
\usepackage{color}
\usepackage{comment}
\usepackage{epstopdf}
\usepackage{cleveref}
\usepackage[svgnames]{xcolor}
\usepackage{enumerate}
\usepackage{braket}
\hypersetup{hidelinks,colorlinks=true,allcolors=DarkBlue}

\theoremstyle{remark}

\begin{document}

\preprint{APS/123-QED}

\title{Optical probing in a bilayer dark-bright condensate system}

\author{N.A. Asriyan}
\affiliation{Department of Physics, M.V. Lomonosov Moscow State University, Moscow 119992, Russia}
\affiliation{N.L. Dukhov Research Institute of Automatics (VNIIA), Moscow 127055, Russia}

\author{I.L. Kurbakov}
\affiliation{Institute for Spectroscopy RAS, Troitsk 108840, Moscow, Russia}

\author{A.K. Fedorov}
\affiliation{Russian Quantum Center, Skolkovo, Moscow 143025, Russia}

\author{Yu.E. Lozovik}
\affiliation{Institute for Spectroscopy RAS, Troitsk 108840, Moscow, Russia}
\affiliation{MIEM, National Research University Higher School of Economics, Moscow 101000, Russia}

\date{\today}
\begin{abstract}
We consider a bilayer system of two-dimensional Bose-Einstein-condensed dipolar dark excitons (upper layer) and bright ones (bottom layer). 
We demonstrate that the interlayer interaction leads to a mixing between excitations from different layers. 
This mixing leads to the appearance of a second spectral branch in the spectrum of bright condensate.
The excitation spectrum of the condensate of dark dipolar excitons then becomes optically accessible during luminescence spectra measurements of the bright condensate,  which allows one to probe its kinetic properties.
This approach is relevant for experimental setups, where detection via conventional techniques remains challenging, in particular, the suggested method is useful for studying dark dipolar excitons in transition metal dichalcogenide monolayers.
\end{abstract}

\maketitle

\section{Introduction}

Achieving full control over quantum many-body systems is of significant importance in both fundamental science and possible applications~\cite{Lukin2014}.
Remarkable progress in experiments with atomic and molecular gases in degenerate regimes makes 
them a perfect playground for revealing novel phases and many-body states of ultracold matter~\cite{Pitaevskii1999,Pitaevskii2008,Bloch2008} 
and also realizing quantum technologies~\cite{Bloch2008,Bloch2012,Bloch2017}.
Another promising platform in this context is many-body systems of quasiparticles, such as excitons in solid-state materials~\cite{Ginzburg1968,Keldysh1968,Lozovik1973,Lozovik2018}. 
Creating excitons with spatially separated electrons and holes increases their lifetime significantly~\cite{Lozovik1973,Lozovik1975,Lozovik2018,shevchenko,Neilson1998,Combescot2008},
which is quite favorable for investigating their collective properties at sufficiently high temperatures~\cite{Moskalenko2000,Butov2012,Butov20122,Fedorov2014,Rapaport2014,T}.
Coupled quantum wells separated by a barrier~\cite{Snoke2010,Butov2012,Butov20122}
and single quantum wells in electric fields~\cite{Dietsche2006,Dietsche20062,Filinov2006} with long-lived two-dimensional (2D) excitons have been intensively studied experimentally (see Ref.~\cite{Combescot2008}). 
Recently, significant attention has been paid to excitons in graphene monolayers separated by an insulating barrier~\cite{LozovikSokolik2008,MacDonald2008,LozovikSokolik2012,Neilson2013},
thin films of topological insulators~\cite{Moore2009,Moore2011,Wang2011,MacDonald2011,EfimkinLozovikSokolik2012,EfimkinLozovik2013},
and  transition metal dichalcogenide (TMDC) monolayers~\cite{Geim2013,Novoselov2014,Rivera2015,Geim2016,Wurstbauer2017,Geim2018}.

In the systems mentioned above, excitons can be optically either bright or dark. 
Properties of bright-exciton condensates, such as condensate density and coherence time, can be directly obtained from emission pattern (luminescence) measurements~\cite{Butov2012},
whereas the detection of dark excitons with existing optical techniques remains challenging.
Understanding properties of dark excitons is of fundamental importance for a wide variety of processes in semiconductors. 
This is also crucial for potential applications such as light harvesting~\cite{Lozovik1985,Butov1999,Bode2009} and quantum-state engineering~\cite{Gershoni2016}. 
Recent proposals for the detection of dark excitons include using near-field coupling to surface plasmon polaritons~\cite{Zhou2017} and monitoring their interactions with a polariton mode~\cite{Kavokin2018}.

\begin{figure}
\includegraphics[width=0.45\textwidth]{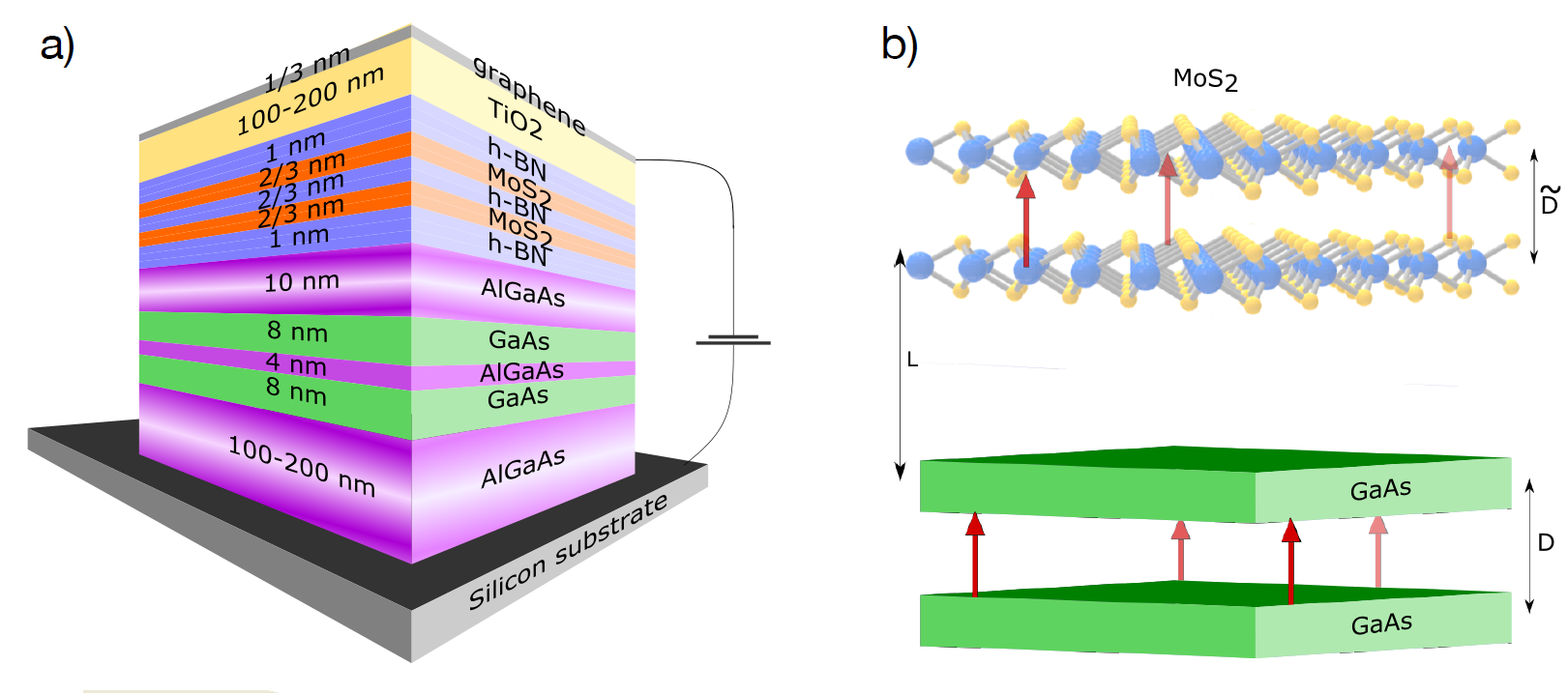}
\vskip -3mm
\caption{Bilayer system of Bose-Einstein-condensed excitons under consideration. 
(a) The suggested experimental realization of the bilayer system of Bose-Einstein-condensed dark (upper layer) excitons in the MoS$_2$ layer and bright excitons (bottom layer) in the GaAs layer with electrons and holes being separated in each layer.
(b) The effective model of the system corresponds to a bilayer system of dipolar particles separated by a distance, $L$.}
\label{fig1}
\end{figure}

In this paper, we study a system of 2D Bose-Einstein-condensed dipolar dark (upper layer) and bright (bottom layer) excitons in a bilayer geometry where the excitons are oriented perpendicularly to the layers (see Fig.~\ref{fig1}).
We demonstrate that this setup offers a possibility of direct probing kinetic properties of the dark-exciton condensate via luminescence spectra measurements of the bright condensate.
This is feasible since the interlayer interaction leads to the appearance of a second spectral branch in the spectrum of bright condensate.
This allows measurement of excitation spectrum and kinetic properties of the dark condensate with the use of conventional luminescence spectra measurements.
Recent experiments on studying excitons in TMDC monolayers~\cite{Geim2013,Novoselov2014,Rivera2015,Geim2016,Wurstbauer2017,Geim2018,Zhou2017} could be accomplished by this method. 
We also note that bilayer systems of dipolar particles have been widely studied 
in the context of ultracold quantum gases~\cite{Shlyapnikov2010,Wang2007,Santos2010,RossiDasSarma2010,Baranov2011,Santos2011,Zinner2012,Fedorov2016}, 
including interesting findings on interlayer superfluidity~\cite{Shlyapnikov2010, Fedorov2016}.

The paper is organized as follows. 
In Sec.~\ref{sec:system}, we describe the system of bilayer dark-bright condensates.
In Sec.~\ref{sec:interlayer}, we reveal the impact of the interlayer interaction between excitons and calculate the excitation spectrum. 
In Sec.~\ref{sec:experiment}, we discuss experimental conditions for optical probing of the excitation spectrum of the condensate of dark dipolar excitons. 
We give our conclusion in Sec.~\ref{sec:conclusion}.

\section{Bilayer dark-bright condensate system}\label{sec:system}

We study a 2D system of Bose-Einstein-condensed dipolar bright and dark excitons in a bilayer geometry with all dipoles oriented in the same direction. 
The Hamiltonian, $\hat{\mathcal{H}}=\hat{\mathcal{H}}_{\rm b}+\hat{\mathcal{H}}_{\rm d}+\hat{\mathcal{H}}_{\rm int}$, includes terms of the following form:
\begin{eqnarray}\label{eq:Hamiltonian}
	&&\!\!\!\hat{\mathcal{H}}_{\rm b (d)}{=}\int\!\hat{\psi}^{+}_{\rm b(d)}({\bf{r}})\left(-\dfrac{\hbar^2}{2m_{\rm d(b)}}\Delta-\mu_{\rm b(d)}\right)\!\hat{\psi}_{\rm b (d)}({\bf{r}})d{\bf{r}} \\
	&&+\int\hat{\psi}^{+}_{\rm b(d)}({\bf{r}})\hat{\psi}^{+}_{\rm b(d)}({\bf{s}})U_{\rm b(d)}({\bf{r}}-{\bf{r}}')\hat{\psi}_{\rm b(d)}({\bf{s}})\hat{\psi}_{\rm b(d)}({\bf{r}})\dfrac{d{\bf{r}}d{\bf{r}}'}{2}, \nonumber
\end{eqnarray}
Here $\hat{\psi}({\bf{r}})$ is the exciton bosonic field operator, $m$ is the exciton mass, 
the subindex ``${\rm b}$'' indicates the bright-exciton condensate, ``${\rm d}$'' indicates the dark-exciton condensate,
and ${\bf{r}}=\{x,y\}$ is the 2D position vector in the layer.
We consider excitons as bosons since the overlap integral of exciton wavefunctions is exponentially small (see Ref.~\cite{Lozovik2017}). 
We also note that chemical potentials $\mu_{\rm b(d)}$ may not be equal to each other since we assume no interlayer hopping.
For bright layers (GaAs) we use a Coulomb-like interaction potential. 
The calculation of the in-layer interaction potential for MoS$_2$ layers is different from the standard treatment due to a specific response in monolayers (see Appendix A and Appendix B).

\begin{figure}
\includegraphics[width=0.5\textwidth]{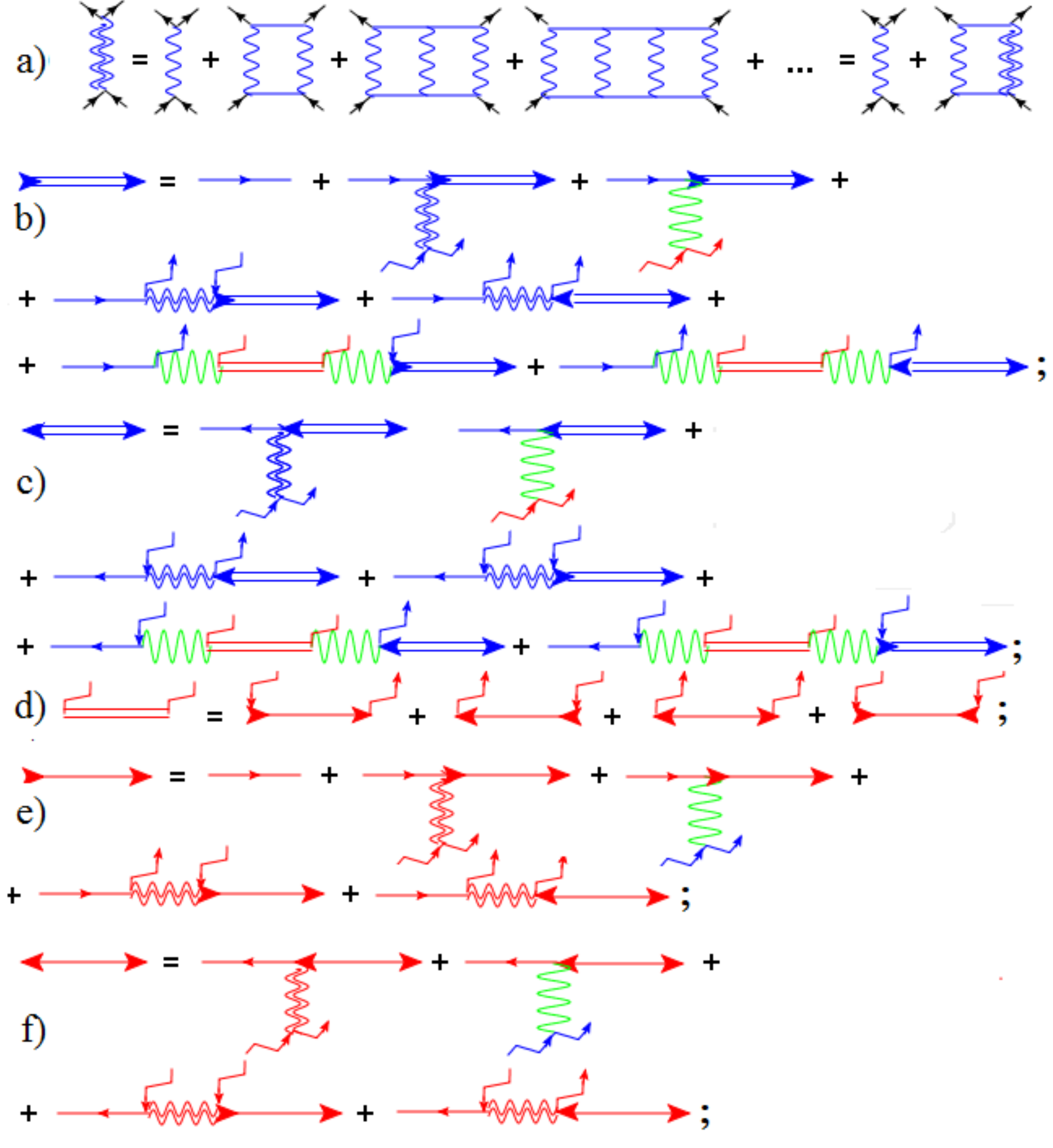}
\vskip -4mm
\caption{Relevant diagrams for the system in the weakly interacting regime.
(a)  Summation of the ladder diagrams.
(b) and (c)  Summation of the diagrams for the normal $G_{\bf k}(\omega)$ and anomalous $F_{\bf k}(\omega)$ Green's functions in the bright layer.
(d) Dynamical structural factor in the dark layer.
(e) and (f) Summation of the diagrams for the normal $\breve G_{\bf k}(\omega)$ and anomalous $\breve F_{\bf k}(\omega)$ Green's functions in the dark layer excluding bare Green's functions of the bright layer exciton's (blue solid lines). 
Blue (red) thin solid lines correspond to bare Green's function of bright (dark) layer. 
Blue (red) wavy lines correspond to the dressed interaction between excitons $U_{\rm b}({\bf k})$ ($U_{\rm d}({\bf k})$) in the bright (dark) layer, whereas the green line stands for the bare interlayer interaction $V({\bf k})$.}
\vskip -6mm
\label{fig2}
\end{figure}

In the dilute regime, in the first Born approximation for both bright and dark layers the in-layer interaction is as follows: 
\begin{equation}\label{eq:interaction1}
	U_{\rm b(d)}({\bf k})-g_{\rm b(d)}=U_{\rm b(d)}^{0}({\bf{k}})-U_{\rm b(d)}^{0}({0}),
\end{equation}
where $g_{\rm b (d)}$ is the coupling constant of excitons in the bright (dark) layer, and ${\bf{k}}$ is the 2D momentum.
However, we cannot use $g_{\rm b(d)}=\int U^0_{\rm b (d)}({\bf{r}})d{\bf{r}}$ explicitly because of the divergency of the interaction potential of rigid dipoles at $r \to 0$.
Therefore, for the weakly correlated system, we dress the bare interaction by the ladder diagrams~\cite{LY2} [see also Fig.~\ref{fig2}(a)]. 
The quantitative approach for calculating $g_{\rm b(d)}$ is the same as that in Ref.~\cite{Voronova2018}. 
To distinguish dressed and bare interactions we use ``0" a superscript below.

Assuming no interlayer hopping, the interaction part of the Hamiltonian is as follows:
\begin{equation}
	\!\!\!\!\!\!\!\hat{\mathcal{H}}_{\rm int}
	{=}\int\hat{\psi}^{+}_{\rm b}({\bf{r}}_{\rm b})\hat{\psi}^{}_{\rm b}({\bf{r}}_{\rm b})V({\bf{r}}_{\rm b}{-}{\bf{r}}_{\rm d})\hat{\psi}^{+}_{\rm d}({\bf{r}}_{\rm d})\hat{\psi}^{}_{\rm d}({\bf{r}_{\rm d}})d{\bf{r}_{\rm b}}d{\bf{r}_{\rm d}}.
\end{equation}

The interlayer interaction may be still taken in the Coulomb form since the interlayer separation $L$ is much larger than the effective screening length in the MoS$_2$ monolayer:
\begin{eqnarray}\label{eq:interaction2}
	&&\!\!V({\bf{R}}){=}\frac{e^2}{\varepsilon}\left(-\frac{1}{\sqrt{R^2+L^2}}-\frac{1}{\sqrt{R^2+(L+D_{\rm b}+{D}_{\rm d})^2}}+\right. \nonumber \\
	&&\left.+\frac{1}{\sqrt{R^2+(L+D_{\rm b})^2}}+\frac{1}{\sqrt{R^2+(L+{D}_{\rm d})^2}}\right).
\end{eqnarray}
where $L$ is the interlayer spacing, 
$D_{\rm b(d)}$ is the effective electron-hole separation in bright(dark) layer, 
$e$ is the electron charge, and 
$\epsilon$ is the interlayer dielectric constant.
In experimentally relevant situations, it is assumed to be equal to the dielectric constant in a bright layer (GaAs). 
This is because the dark layer (TMDC) is very thin and lies on a thicker structure of GaAs quantum well.

In the first Born approximation the interaction potential~(\ref{eq:interaction2}) satisfies the relation $h=\int V({\bf{R}})d{\bf{R}}=0$,
which complicates the ordinary method of finding a bound state in 2D potentials finite at the origin~\cite{LL3,Simon1976}. 
If the bound state energy is significant then one can expect the formation of interlayer exciton biexcitons (dimers). 
However, in the considered experimental setup, we expect that the bound state energy is small so the biexciton (dimer) physics is still not important. 
We also note that the effects of the formation of bound states in bilayer dipolar systems have been studied~\cite{Lozovik1975,Shlyapnikov2010,Santos2010,Yudson1997}.

\section{Excitation spectra}\label{sec:interlayer}

The key of our work is to reveal the impact of the interlayer interaction in the luminescence spectrum of the optically accessible bright condensate.
Assuming the system to be homogeneous, it can be expressed in the following form (see, e.g., Ref.~\cite{Citrin1993}):
\begin{equation}\label{eq:luminescent}
	I_{\rm b}(\varphi,\theta,\omega_{\rm phot})\propto\frac{E_g}{2\pi\tau}iG_{\bf k}^+(\omega).
\end{equation}
Here $\tau$ is the bright condensate lifetime, 
$\omega_{\rm phot}=\omega+(E_g+\mu)/\hbar$ is the luminescence frequency,
$E_g$ is the exciton gap,
$\mu$ is the chemical potential, and 
the exciton momenta ${\bf k}$ depends on polar and azimuthal angles in the free space as ${\bf k}=\hbar(E_g/c_0)\sin\theta\{\cos\varphi,\sin\varphi\}$, where $c_0$ is speed of light in a vacuum.
The correlation function in the Keldysh form in Eq.~(\ref{eq:luminescent}) is as follows:
\begin{equation}\label{eq:Gp+_definition}
	G_{\bf{k}}^+(\omega)\equiv-i\int_{-\infty}^{\infty} dt e^{i\omega t} \langle \hat a_{\bf{k}}^+(0)\hat a^{}_{\bf{k}}(t) \rangle,
\end{equation}
where $\hat a_{\bf{k}}(t) =  e^{i\mathcal{H}t/\hbar}\hat a_{\bf{k}} e^{-i\mathcal{H}t/\hbar}$ is the bright layer exciton annihilation operator in the Heisenberg representation.

\begin{figure}
    \centering
    \includegraphics[width=0.45\textwidth]{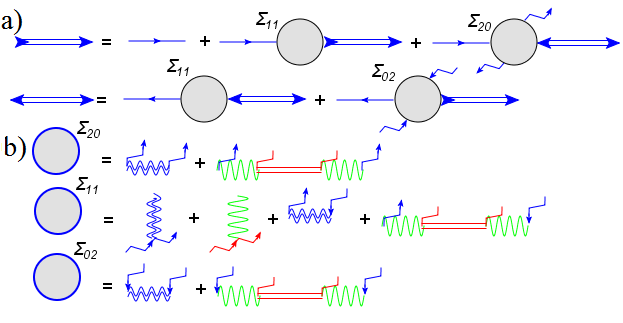}
    \vskip-4mm
    \caption{(a) Expansion of Green's functions in terms of self-energy terms in a generic case of a BEC system. 
    Here the first and the second indices indicate the number of outgoing and incoming condensate lines, correspondingly. 
    (b) Explicit expressions for self-energy terms of the bilayer system}
    \label{blocks1}
\end{figure}

In order to obtain the luminescence spectrum, we calculate expression (\ref{eq:Gp+_definition}) using the Green's functions approach in the dilute regime at $T=0$, where one can assume the negligibility of the loop diagrams.
We also consider the system of excitons having a single spin degree of freedom. 
This is because in the case of multiple spin branches, at $T=0$ both condensate and non-condensate occupy the lowest spin branch only (see Appendix C). 
We suppose that the lowest spin branch of the bright layer is bright as well as in Ref.~\cite{Butov2012}.
All the relevant diagrams describing the system are presented in Fig.~\ref{fig2}. 
Here and further we use a tilde instead of b (d) subindex for highlighting quantities related to the dark condensate, 
because we would like to emphasize that the following calculations are quite general till Eq. (\ref{operator_transformation}) and all the results may be formulated for the second layer swapping tilde expressions with the ones without tildes.  

\begin{widetext}
In the dilute regime for a Bose-Einstein condensate (BEC) system normal $G_{\bf{k}}(\omega)$ and anomalous $F_{\bf{k}}(\omega)$ Green's functions can be obtained from the standard Belyaev system~\cite{Belyaev}, which can be presented as follows:
\begin{equation}\label{Belyaev_cases}
	\begin{cases}
	G_{\bf{k}}(\omega)=&G^0_{\bf{k}}(\omega)+G^0_{\bf{k}}(\omega)\Sigma^{11}_{\bf{k}}(\omega)G_{\bf{k}}(\omega)+G^0_{\bf{k}}(\omega)\Sigma^{20}_{\bf{k}}(\omega)F_{\bf{k}}(\omega),\\
	F_{\bf{k}}(\omega)=&G^0_{-\bf{k}}(-\omega)\Sigma^{11}_{\bf{k}}(\omega)F_{\bf{k}}(\omega)+G^0_{-\bf{k}}(-\omega)\Sigma^{02}_{\bf{k}}(\omega)G_{\bf{k}}(\omega).\\
\end{cases}
\end{equation}
Using the explicit expression for the Green's function of free boson $G^{(0)}_{\bf k}(\omega)={\hbar}/\left({\hbar\omega-T_{\bf k}+\mu+i\delta}\right)$,
where $T_{{\bf k}}={\hbar^2 k^2}/({2m})$, we express the formal solution of the Belyaev system (\ref{Belyaev_cases}) in the following form:
\begin{equation}
\label{solution}
\begin{cases}
    \displaystyle{G_{\bf{k}}(\omega)= \frac{\hbar\omega+T_{\bf{k}}+S_{\bf{k}}(\omega)-A_{\bf{k}}(\omega)-\mu}{(\hbar\omega-A_{\bf{k}}(\omega))^2-(T_{\bf{k}}+S_{\bf{k}}(\omega)-\mu)^2+\Sigma^{20}_{\bf{k}}(\omega)\Sigma^	{02}_{\bf{k}}(\omega)}}\\
    \displaystyle{F_{\bf{k}}(\omega)=-\hbar\frac{\Sigma^{02}_{\bf{k}}(\omega)}
    {(\hbar\omega-A_{\bf{k}}(\omega))^2-(T_{\bf{k}}+S_{\bf{k}}	(\omega)-\mu)^2+\Sigma^{20}_{\bf{k}}(\omega)\Sigma^{02}_{\bf{k}}(\omega)}},
\end{cases}
\end{equation}
where $A_{\bf{k}}(\omega)=\left[{\Sigma^{11}_{\bf{k}}(\omega)-\Sigma^{11}_{-\bf{k}}(-\omega)}\right]/2$ and $S_{\bf{k}}(\omega)=\left[{\Sigma^{11}_{\bf{k}}(\omega)+\Sigma^{11}_{-\bf{k}}(-\omega)}\right]/2$.
Corresponding diagrams are presented in Fig.~\ref{blocks1}(a).
By comparing the diagrams in Fig.~\ref{blocks1}a and Eq.~(\ref{Belyaev_cases}), we find explicit graphical expressions for self-energy terms, which are presented in a diagrammatic form in Fig.~\ref{blocks1}(b).
\end{widetext}

For deriving analytical expressions for self-energy terms one needs to obtain the dynamical structural factor [see Fig \ref{fig2}d], 
which can be expressed using Green's functions $\breve{G}_{\bf{k}}(\omega)$ and $\breve{F}_{\bf{k}}(\omega)$ [see Fig~\ref{fig2}(e) and Fig~\ref{fig2}(f)]. 
Calculating normal $\breve{G}_{\bf{k}}(\omega)$ and anomalous $\breve{F}_{\bf{k}}(\omega)$ Green's functions for the dark layer also requires the use of the Belyaev's system. 
For this we exclude diagrams that contain the bright layer's bare Green's functions (blue solid lines). 
The collections of terms belonging to different self-energy terms [see Fig~\ref{fig2}(e) and Fig~\ref{fig2}(f)] give the following expression:
\begin{equation}\label{single_layer}
\!\!\!\!\!\!
\begin{cases}
	\displaystyle{\breve{G}_{\bf{k}}(\omega)=\tilde G^{(0)}_{\bf{k}}(\omega)\left(1+\frac{\tilde{\mu}+\tilde U_{\bf{k}}}{\hbar}\breve{G}_{\bf{k}}(\omega)+\frac{\tilde U_{\bf{k}}}{\hbar}\breve{F}_{\bf{k}}(\omega)\right),}\\
	\displaystyle{\breve{F}_{\bf{k}}(\omega)=\tilde G^{(0)}_{\bf{-k}}(-\omega)\left(\frac{\tilde{\mu}+\tilde U_{\bf{k}}}{\hbar}\breve{F}_{\bf{k}}(\omega)+\frac{\tilde U_{\bf{k}}}{\hbar}\breve{G}_{\bf{k}}(\omega)\right),}
\end{cases}
\end{equation}
where we use use the following notations: $U_{\bf{k}} = n_0 U(\bf{k})$, $\tilde U_{\bf{k}}=\tilde n_0 U(\bf{k})$,$V_{\bf{k}} = n_0 V(\bf{k})$, $\tilde V_{\bf{k}}=\tilde n_0 V({\bf{k}})$, 
where $V({\bf{k}})=\int V({\bf{r}}) \rm exp(-i\bf{kr}) d \bf{r}$, and $n_0$ is the condensate density, and $\tilde \mu = g\tilde n_0 + hn_0$ is the chemical potential of the dark layer.

By solving these equations and substituting $\breve{G}_{\bf{k}}(\omega)$ and $\breve{F}_{\bf{k}}(\omega)$ into the expression for the structural factor [Fig~\ref{fig2})(d)] we obtain the solution of the Belyaev system
(\ref{Belyaev_cases}) in the following form:
\begin{equation}
\!\!\!\!
\begin{cases}
	\displaystyle{G_{\bf k}(\omega)=
	\hbar\frac{\hbar\omega+T_{\bf k}+\Sigma_{\bf k}(\omega)}{\hbar^2\omega^2-
	\left[T_{\bf k}+\Sigma_{\bf k}(\omega)\right]^2+\Sigma_{\bf k}(\omega)^2}}, \\
		\displaystyle{F_{\bf k}(\omega)=-\hbar\frac{\Sigma_{\bf k}(\omega)}{\hbar^2\omega^2-
	\left[T_{\bf k}+\Sigma_{\bf k}(\omega)\right]^2+\Sigma_{\bf k}(\omega)^2}},
\end{cases}
\end{equation}
with the following notation for the self-energy:
\begin{equation}\label{sigma_1}
        \Sigma_{{\bf k}}(\omega)\equiv U_{{\bf k}}+\frac{2V_{{\bf k}}\tilde V_{{\bf k}}\tilde{T}_{{\bf k}}}{\hbar^2\omega^2-\tilde{T}_{{\bf k}}\left(\tilde{T}_{{\bf k}}+2\tilde U_{{\bf k}}\right)+i\delta}.
\end{equation}

By expanding these expressions in the form of simple fractions, we obtain
\begin{equation}\label{f_fraction}
\begin{split}
	F_{\bf{k}}(\omega)=-\frac{\hbar u_{{\bf k}}^+v_{{\bf k}}^+}{\hbar\omega-\varepsilon_{\bf k}^++i\delta}+\frac{\hbar u_{{\bf k}}^+v_{{\bf k}}^+}{\hbar\omega+\varepsilon_{\bf k}^+-i\delta}\\
	-\frac{\hbar u_{{\bf k}}^-v_{{\bf k}}^-}{\hbar\omega-\varepsilon_{\bf k}^-+i\delta}+\frac{\hbar u_{{\bf k}}^-v_{{\bf k}}^-}{\hbar\omega+\varepsilon_{\bf k}^--i\delta},
\end{split}
\end{equation}
\begin{equation}\label{g_fraction}
\begin{split}
	G_{\bf{k}}(\omega)=\frac{\hbar u_{{\bf k}}^+{}^2}{\hbar\omega-\varepsilon_{\bf k}^++i\delta}-\frac{\hbar v_{{\bf k}}^+{}^2}{\hbar\omega+\varepsilon_{\bf k}^+-i\delta} \\
	+\frac{\hbar u_{{\bf k}}^-{}^2}{\hbar\omega-\varepsilon_{\bf k}^-+i\delta}-\frac{\hbar v_{{\bf k}}^-{}^2}{\hbar\omega+\varepsilon_{\bf k}^--i\delta}.
\end{split}
\end{equation}
Here the following notations are introduced:
\begin{eqnarray}\label{u}
	u_{\bf k}^{\pm2}=\frac{(U_{{\bf k}}+T_{{\bf k}}+\varepsilon^{\pm}_{{\bf k}})((\tilde \varepsilon_{{\bf k}}^0)^2-\varepsilon^{\pm}_{{\bf k}}{}^2)-2\tilde T_{{\bf k}}V_{{\bf k}}\tilde V_{{\bf k}}}{2\varepsilon^{\pm}_{{\bf k}}(\varepsilon_{\bf k}^{\mp}{}^2-\varepsilon_{\bf k}^{\pm}{}^2)},
\end{eqnarray}
\begin{equation}\label{v}
	v^{\pm2}_{{\bf k}}=\frac{(U_{{\bf k}}+T_{{\bf k}}-\varepsilon^{\pm}_{{\bf k}})((\tilde \varepsilon_{{\bf k}}^0)^2-\varepsilon^{\pm}_{{\bf k}}{}^2)-2\tilde T_{{\bf k}}V_{{\bf k}}\tilde V_{{\bf k}}}{2\varepsilon^{\pm}_{{\bf k}}(\varepsilon_{\bf k}^{\mp}{}^2-\varepsilon_{\bf k}^{\pm}{}^2)},
\end{equation}
\begin{equation}\label{uv}
	u_{{\bf k}}^{\pm}v_{{\bf k}}^{\pm}=\frac{U_{{\bf k}}((\tilde \varepsilon_{{\bf k}}^0)^2-\varepsilon^{\pm}_{{\bf k}}{}^2)-2\tilde T_{{\bf k}}V_{{\bf k}}\tilde V_{{\bf k}}}{2\varepsilon^{\pm}_{{\bf k}}(\varepsilon_{\bf k}^{\mp}{}^2-\varepsilon_{\bf k}^{\pm}{}^2)}.
\end{equation}
\begin{equation}\label{solution}
\varepsilon_{\bf{k}}^{\pm}=\sqrt{\frac{(\varepsilon_{\bf{k}}^0)^2+(\tilde \varepsilon_{\bf{k}}^0)^2}{2}\pm|\Delta_{\bf{k}}|},
\end{equation}
\begin{equation}\label{diskriminant}
    \Delta_{\bf{k}}^2=\frac{((\varepsilon_{\bf{k}}^0)^2-(\tilde \varepsilon_{\bf{k}}^0)^2)^2}{4}+4T_{\bf{k}}\tilde T_{\bf{k}}V_{\bf{k}}\tilde V_{\bf{k}},
\end{equation}
where $(\varepsilon_{\bf{k}}^0)^2=T_{\bf{k}}^2+2T_{\bf{k}}U_{\bf{k}}$ and $(\tilde \varepsilon_{\bf{k}}^0)^2=\tilde T_{\bf{k}}^2+2\tilde T_{\bf{k}}\tilde U_{\bf{k}}$ are excitation spectra without interlayer interaction, 
and $\varepsilon_{\bf{k}}^{\pm}$ are the ones with the interaction.

For the expansions presented above to be valid the stability of the homogeneous system is required, i.e., $(\varepsilon_{\bf{k}}^0)^2>0$, $(\tilde \varepsilon_{\bf{k}}^0)^2>0$, and $ (\varepsilon^{\pm}_{\bf{k}})^2>0$ (for details, see Appendix D).

The obtained Green's functions given by Eqs.~(\ref{f_fraction})-(\ref{g_fraction}) correspond {\it exactly} to the the originally considered bilayer system with the following diagonalized Hamiltonian: 
\begin{equation}\label{diagonalized}
	\hat{\mathcal{H}}=\sum_{{{\bf k}\neq 0}} \varepsilon_{\bf k}^+\hat{\alpha}^+_{{\bf k}}\hat{\alpha}_{{\bf k}}+\varepsilon_{\bf k}^-\hat{\beta}^+_{{\bf k}}\hat{\beta}_{{\bf k}}.
\end{equation}
Here the new annihilation operators may be expressed in terms of exciton annihilation operators $\hat a_{{\bf k}}$ and $\hat b_{{\bf k}}$ for the bright and light layers correspondingly, from the following equations:

\begin{equation}
\label{operator_transformation}
\begin{cases}
    \hat a_{{\bf k}}=u^+_{{\bf k}}\hat{\alpha}_{{\bf k}}-v^+_{{\bf k}}\hat{\alpha}^+_{\bf{-k}}+u^-_{{\bf k}}\hat{\beta}_{{\bf k}}-v^-_{{\bf k}}\hat{\beta}^+_{\bf{-k}},\\
    \hat b_{{\bf k}}=\tilde u^+_{{\bf k}}\hat{\alpha}_{{\bf k}}-\tilde v^+_{{\bf k}}\hat{\alpha}^+_{\bf{-k}}+\tilde u^-_{{\bf k}}\hat{\beta}_{{\bf k}}-\tilde v^-_{{\bf k}}\hat{\beta}^+_{\bf{-k}}.
\end{cases}
\end{equation}
To obtain expressions for $\tilde u^{\pm}_{{\bf k}}$ and $\tilde v^{\pm}_{{\bf k}}$, we use Eqs.~(\ref{u}) and (\ref{v}), swapping all tilde expressions with the ones without tildes.
By substituting these operators in the expression for the Keldysh correlation function (\ref{eq:Gp+_definition}), we obtain the following result:
\begin{equation}\label{eq:result}
	\!\!iG_{{\bf k}}^+(\omega)=2\pi\hbar\left(v_{{\bf k}}^+{}^2\delta(\hbar\omega+\varepsilon_{{\bf k}}^+)+v_{{\bf k}}^-{}^2\delta(\hbar\omega+\varepsilon_{{\bf k}}^-)\right),
\end{equation}
where $n_{{\bf k}}=v_{{\bf k}}^{+\,2}+v_{{\bf k}}^{-\,2}$.

\begin{figure}
\includegraphics[width=0.45\textwidth]{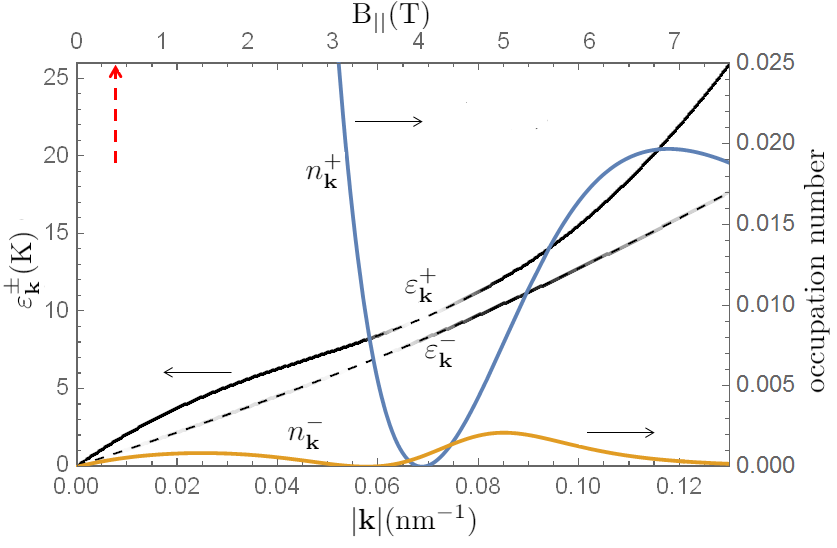}
\vskip -4mm
\caption{Excitation spectra as a function of the momentum are shown by dashed lines (left scale). 
The momentum dependence of the occupation numbers on these branches is highlighted in color and shown by solid lines (see right scale). 
The occupation is also indicated by gray filling of dashed lines. 
Two areas of coexistence of occupation in both branches are clearly visible. 
The upper scale shows the magnitude of the magnetic field needed for observation of excitation spectra along the normal surface ($\theta=0$). 
The red arrow indicates the boundary of the radiation zone in free space.}
\label{fig3}
\end{figure}

By analyzing expressions (\ref{v}), (\ref{solution}), and (\ref{eq:result}), we conclude that in the case of the absence of the interlayer interaction one has two spectral branches, but only one branch has non-zero population, i.e., it is optically accessible. 
Taking into account the interlayer interaction leads to the significantly large occupation of the second branch in certain momentum regions.
Therefore, by inclusion of the interlayer interaction the excitation spectrum of the dark condensate becomes optically accessible during luminescence spectra measurements of the bright condensate, which allows one to probe its kinetic properties.
Figure~\ref{fig3} clearly demonstrates two areas in which the effect of mixing excitations leads to a noticeable population of both branches near the certain momentum region.
The latter would indicate the existence of a large-scale coherence in both layers.
We note that due to the in-plane momentum conservation the value $k$ is bounded by $\hbar E_g/c_0$ (the radiation zone in free space). 
This is because excitation energies are much smaller than $E_g=1.57 eV$ in GaAs. 
Thus, in an experimental setup, it may be useful to apply the in-plane magnetic field to shift the dispersion curve of excitations~\cite{Bparallel}. 
Magnetic fields needed to observe the excitation spectrum along the normal surface (at $\theta$=0) are indicated in Fig.~\ref{fig3}.

\section{Experimental realization}\label{sec:experiment}

We assume that the suggested approach for optical probing in a bilayer dark-bright condensate system is relevant for studying excitons in TMDC monolayers. 
In particular, we consider a bilayer structure with bright excitons in a GaAs/AlGaAs/GaAs layer and dark excitons in a MoS$_2$/hBN/MoS$_2$ structure.
For numerical estimation we use parameters from Table \ref{tab:1} and assume interlayer separation of $L=15 \rm\text{ } nm$.

\begin{table}[H]
	\begin{center}
	\begin{tabular}{|ccccccc|}
	\hline
	&$n_0, 10^{10} {\rm cm}^{-2}$&$m/m_0$&$m_{\rm r}/m_0$&$D$, nm&$\epsilon$&$\alpha_{2D},\AA$\\
	\hline
	TMDC&15&1&0.25&1.333&8.7&6.60\\
	GaAs&1&0.22&0.0467&12&12.5&$-$\\
	\hline
	\end{tabular}
	\end{center}
	\vskip-4mm
	\caption{Here $n_0$ is the condensate density, 
	$m/m_0$ and $m_{\rm r}/m_0$ are ratios of exciton mass and reduced mass to free electron mass, 
	$D$ is the effective electron-hole separation within the single layer, 
	and $\epsilon$ is the dielectric constant of the surrounding medium (see Appendix B). 
	The interlayer dielectric constant  is assumed to be equal to the one in GaAs (see text). 
	$\alpha_{2D}$ is the 2D polarizability of the MoS$_2$ monolayer~\cite{Berkelbach2013}.}
	\label{tab:1}
\end{table}

We would like to note that there is a huge inhomogeneous broadening of the exciton resonance in TDMC structures.
The advantage of our method is that it is sensitive to the inhomogeneous broadening in a bright layer in the GaAs structure, which is very small, but not to the inhomogeneous broadening in the dark layer in TDMC structures. 
We also note that if the interlayer hopping can be neglected, effects of excitation mixing and splitting spectral branches take place if both layers are in the BEC phase~\cite{Comment}. 
If there is no condensate in the dark layer, then there is no occupation of the second spectral branch~\cite{ssc144000399}. 
We expect that the suggested approach is relevant for experimental setups, where detection via conventional techniques remains challenging, 
in particular, the considered method is useful for studying excitons in TMDC monolayers.

To take into account finite coherence lengths $\xi_b$ ($\xi_d$) and coherence times $\tau_b$ ($\tau_d$),
the excitation spectrum  should be treated by considering uncertainties in momentum and energy of order $\max(\hbar/\xi_b,\hbar/\xi_d)$ and  $\max(\hbar/\tau_b,\hbar/\tau_d)$, correspondingly.

\section{Conclusion}\label{sec:conclusion} 

In the present work, we have considered the system of bilayer dark-bright condensates.
We have revealed the impact of the interlayer interaction between excitons and calculated the excitation spectrum. 
We have demonstrated that the excitation spectrum of the condensate of dark dipolar excitons becomes accessible for optical probing under realistic experimental conditions. 

We would like to emphasize, that our approach is still applicable in the finite temperature case. 
Neglecting the loop diagrams is still valid up to sufficiently high temperatures of the same order of magnitude as the Berezinskii-Kosterlitz-Thouless (BKT) crossover temperature [see Ref.~\cite{Voronova2018} Fig.~3(a)]. 
However, a quantitative analysis is beyond the scope of the current paper.

\section*{Acknowledgement}

A.K.F. is supported by the RFBR grant (17-08-00742).
I.L.K. is supported by the RFBR grant (17-02-01322).
Y.E.L. is supported by program of Higher School of Economics.
*Corresponding author: {\href{mailto:lozovik@isan.troitsk.ru}{lozovik@isan.troitsk.ru}}.

\setcounter{equation}{0}
\renewcommand{\theequation}{A\arabic{equation}}

\section*{Appendix A: Keldysh-Rytova potential for a bilayer system}

Here we consider a bilayer system separated by a distance $D$ with 2D polarizabilities of single monolayers $\alpha_1$ and $\alpha_2$, 
which are embedded into a dielectric medium with the dielectric constant $\epsilon$ and a point charge $e$, located in the first layer at the origin.
In order to find the electrostatic potential in the whole space, we start from the Poisson's equation of the following form (in the way it is done in Ref.~\cite{Cudazzo2011}):
\begin{equation}
	\Delta \varphi=-4\pi n({\bf{r}}_{\rm 3D}),
\end{equation}
where $n$ is the total charge density, which is a sum $n=n_{\rm ext}+n_{\rm ind}$ of external (point charge) and induced densities.
Then $n_{\rm ind}$ has three contributions of the following form:
\begin{equation}
\begin{split}
	n_{\rm ind}=&-{\rm div}{{\bf{P}}}=\frac{\epsilon-1}{4\pi}\Delta \varphi(r, z)+\\ 
	&+\delta(z) \alpha_1 \Delta_r \varphi(r, z=0)+ \\
	&+\delta(z-D) \alpha_2 \Delta_r \varphi(r, z=D).
\end{split}
\end{equation}
The first one corresponds to the polarization of the dielectric environment, whereas the remaining ones are charges confined in monolayers. 
Here we use the following notation: ${\bf{r}}_{3D}=r{\bf{e}}_r+z{\bf{e}}_z$, where $r$ is the in-plane coordinate and $z$ is the normal coordinate.
We then arrive at the following equation:
\begin{equation}\label{Poisson}
\begin{split}
	\epsilon\Delta \varphi=&-4\pi[e\delta({\bf{r}}_{3D})+\delta(z) \alpha_1 \Delta_r \varphi({\bf{r}}, z=0)+\\ 
	&+\delta(z-D) \alpha_2 \Delta_r \varphi({\bf{r}}, z=D)].
\end{split}
\end{equation}
One may see that the impact of the dielectric environment may be replaced by reducing $e$, $\alpha_1$, and $\alpha_2$. 
Thus it is possible to find the potential for the vacuum case first, and then reduce these values in $\epsilon$ times.

Applying the Fourier transform to Eq.~(\ref{Poisson}), we obtain:
\begin{equation}\label{Fourier poisson}
\begin{split}
	&\!\!\!\!\!(k_{z}^2+k^2)\varphi(k_z, {\bf{k}}){+}2\alpha_2 k^2 e^{-ikD}{\int}dk_z' \varphi(k_z', {\bf{k}})e^{ik_z'D} \\
	&\!\!=4\pi e-2\alpha_1 k^2\int dk_z' \varphi(k_z', {\bf{k}}),
\end{split}
\end{equation}
where ${\bf{k}}$ and $k_z$ are in-plane and normal components of the wave vector, respectively. 
Here and below we use the following expressions for the screening lengths: $\rho_1=2\pi \alpha_1$ and $\rho_2=2\pi \alpha_2$.

From Eq.~(\ref{Fourier poisson}) one can conclude that the potential has the following form:
\begin{equation}\label{guess}
	\varphi(k_z, |{\bf{k}}|)=\dfrac{c_1(|{\bf{k}}|)e^{-ik_zD}+c_2(|{\bf{k}}|)}{k_z^2+k^2}.
\end{equation}
Substituting Eq.~(\ref{guess}) into Eq.~(\ref{Fourier poisson}) and then integrating over $k_z'$, we obtain the following equations for $c_1$ and $c_2$ ($k=|{\bf{k}}|)$:
\begin{equation}
\begin{cases}
	c_1+k\rho_2[c_1+c_2 e^{-kD}]=0,\\
	c_2+k\rho_1[c_1 e^{-kD}+c_2]=4\pi e.
\end{cases}
\end{equation}
By solving these equation, we arrive at the following expression for the electrostatic potential:
\begin{equation}\label{inverse_potential}
\begin{split}
	&\varphi(k_z, k, D)=\dfrac{c_1(k)e^{-ik_zD}+c_2(k)}{k_z^2+k^2}=\\
	&\dfrac{4\pi e}{(k_z^2+k^2)}\dfrac{1+k\rho_2-k\rho_2e^{-kD}e^{-ik_zD}}{(1+k\rho_1)(1+k\rho_2)- k^2\rho_1 \rho_2e^{-2kD}}.
\end{split}
\end{equation}
In order to calculate the interexciton interaction potential, we need only special cases of $z=0$ and $z=D$. 
Taking the inverse Fourier transform with fixed $z$ we obtain the following:
\begin{equation}\label{z0}
	\!\!\!\!\varphi_{2D}(k,D)=\dfrac{2\pi e e^{-kD}}{k((1+k\rho_1)(1+k\rho_2)-k^2\rho_1 \rho_2 e^{-2kD})}
\end{equation}
and
\begin{equation}\label{sensor}
	\!\!\!\!\varphi_{2D}(k,0)=\dfrac{2\pi e\left(1+k\rho_2\left(1-e^{-2kD}\right)\right)}{k((1+k\rho_1)(1+k\rho_2)-k^2\rho_1 \rho_2 e^{-2kD})}.
\end{equation}

We note that Eq.~(\ref{sensor}) is not the same as the Keldysh-Rytova potential~\cite{Keldysh1979,Rytova1967} for the monolayer case since thought $z=0$, here it is calculated in the presence of the second layer. 
However, by setting $D=0$ we obtain the Keldysh-Rytova potential with the effective screening length $\rho_{\rm eff}=\rho_1+\rho_2$.

\setcounter{equation}{0}
\renewcommand{\theequation}{B\arabic{equation}}

\section*{Appendix B: In-layer interaction potentials}

In the case of the bright (GaAs) layer, we use the standard Coulomb potential as follows:
\begin{equation}\label{light_eff}
\begin{split}
	U^0_{\rm b}({\bf{k}})=\dfrac{4\pi e^2D_{\rm b}}{\epsilon_{\rm b}}\left(\dfrac{1-e^{-kD_{\rm b}}}{kD_{\rm b}}\right).
\end{split}
\end{equation}

In order to describe the interaction in the dark (MoS$_2$) layer, we use Eqs.~(\ref{z0}) and (\ref{sensor}) for the system considered on Fig.~\ref{fig1} with the use of several simplifications. 
First, we consider the bilayer system of MoS$_2$ embedded in the dielectric medium with the dielectric constant $\epsilon_{\rm_d}=(\epsilon_{\text{GaAs}}+\epsilon_{\text{TiO}_2})/2$,
which includes neglecting the thickness of the 3hBN/ MoS$_2$/2hBN/MoS$_2$/3hBN system. 
Second, we neglect the thickness of MoS$_2$ monolayers and treat them as two absolutely thin sheets separated by the distance $D_{\rm d}=4/3$ nm.

Reducing charges and screening lengths in $\epsilon_{\rm_d}$ times and using the fact that both layers are the same ($\rho_1=\rho_2=\rho$), we obtain the following expression for the in-layer interexciton interaction:
\begin{equation}\label{exact}
\begin{split}
	U^0_{\rm d}({\bf{k}})=&\frac{4\pi e^2}{\epsilon_{\rm_d} k((1+{k\rho}/{\epsilon_{\rm_d}})^2-({k\rho}/{\epsilon_{\rm_d}})^2e^{-2kD_{\rm d}})} \\ 
	\times&\left(\left(1-e^{-kD_{\rm d}}\right)+\frac{k\rho}{\epsilon_{\rm d}}\left(1-e^{-2kD_{\rm d}}\right)\right).
\end{split}
\end{equation}

Although interactions in bright and dark layers have different functional forms, they have the same divergent behavior at the $r\to0$ limit. 
In order to check it, we integrate the angular part of Eqs.~(\ref{light_eff}) (\ref{exact}) and arrive at the following expression:
\begin{equation}
    U_{\rm b(d)}(r)=\int_0^{\infty} f_{\rm b(d)}(r,k) J_0(kr) dk.
\end{equation}
If $r\to0$, the main contribution arises from $k\approx[0;1/r]$ region due to the oscillatory behaviour of the Bessel function at $kr>>1$. 
So, small $r$ behaviour of the interaction potential is governed by $U(k)$ at small momenta. 
Expanding Eqs.~(\ref{light_eff}) and (\ref{exact}), we get the same asymptotic form:
\begin{equation}\label{asymptotic}
    U^{0}_{\rm b(d)}({\bf{k}})=\frac{4\pi e^2D_{\rm b(d)}}{\epsilon_{\rm b(d)}}-\frac{2\pi e^2D_{\rm b(d)}^2k}{\epsilon_{\rm b(d)}}
\end{equation}
We note that the first term is the bare coupling constant, which has to be dressed.

\setcounter{equation}{0}
\renewcommand{\theequation}{C\arabic{equation}}
\section*{Appendix C: Spin effects}

In the boson limit the spin relaxation of excitons is suppressed~\cite{Holleitner2010}.
Therefore, both condensate and non-condensate occupies the lowest spin branch. 
This justifies the consideration of only one spin component for excitons at $T=0$ in the weakly interacting regime.
However, this is not the case for sufficiently low finite temperatures. 
To prevent the significant thermal occupation of higher spin branches, one may take into account the exchange splitting or use the Zeeman effect. 
Neglecting the loop diagrams is still valid up to sufficiently high temperatures, i.e., temperatures of the same order of magnitude as the BKT crossover temperature.
As we may see in Ref.~\cite{Voronova2018} (Fig. 3), the BKT crossover temperature dependence on a magnetic field is quite weak.
However, a quantitative analysis is beyond the scope of the present research.

\setcounter{equation}{0}
\renewcommand{\theequation}{D\arabic{equation}}
\section*{Appendix D: Homogeneous system stability conditions}

\begin{figure}
\includegraphics[width=0.5\textwidth]{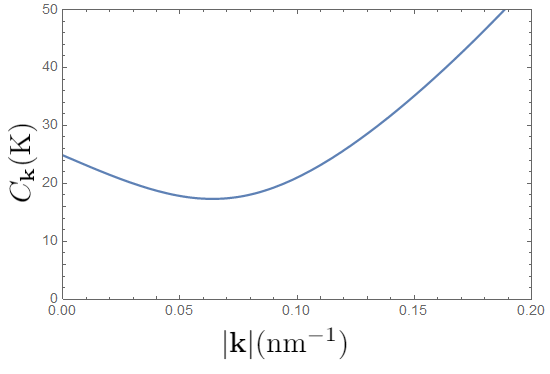}
\vskip -6mm
\caption{$C_k(|{\bf{k}}|)$ is shown as a function of the momenta. 
This value does not cross the $k$-axis, which confirms the stability of the homogeneous system under consideration.} 
\label{fig5}
\vskip -6mm
\end{figure}

Stability conditions in both layers without the interlayer interaction correspond to the following expressions: 
\begin{equation}
	(\varepsilon_{\bf{k}}^0)^2>0, \quad (\tilde \varepsilon_{\bf{k}}^0)^2>0.
\end{equation}
The additional condition for the bilayer case has the following form:
\begin{equation}
    (\varepsilon_{\bf{k}}^{\pm})^2>0.
\end{equation}
From Eq.~(\ref{diskriminant}) one can see that $\Delta_{\bf{k}}^2>0$, since $V_{\bf{k}}<0$ and $\tilde V_{\bf{k}}<0$. 
This provides $(\varepsilon_{\bf{k}}^{+})^2>0$. 
For $(\varepsilon_{\bf{k}}^{-})^2$ to be non-negative, we have
\begin{equation}\label{criteria}
    \frac{\left((\varepsilon_{\bf{k}}^0)^2+(\tilde \varepsilon_{\bf{k}}^0)^2\right)^2}{4}>\Delta_{\bf{k}}^2.
\end{equation}
Expanding Eq.~(\ref{criteria}) we obtain the following expression:
\begin{equation}
    C_{\bf{k}}^2\equiv T_{\bf{k}}\tilde T_{\bf{k}}+2\tilde T_{\bf{k}} U_{\bf{k}}+2 T_{\bf{k}}\tilde U_{\bf{k}}+4 U_{\bf{k}}\tilde U_{\bf{k}}-4V_{\bf{k}}\tilde V_{\bf{k}}>0.
\end{equation}
For the bilayer system under consideration this inequality holds, which can be shown by calculation (see Fig. \ref{fig5}).


\begin{thebibliography}{99}
\bibitem{Comment}
In the diagram language, in the approximation with neglecting loop diagrams these effects are to the impact of the diagrams containing dynamical structural factors, which are nonzero if there is BEC in both layers only.
\end{thebibliography}
\end{document}